\documentclass[aps,reprint,superscriptaddress,showpacs]{revtex4-1}

\usepackage{graphicx}
\usepackage{dcolumn}
\usepackage{bm}
\usepackage{natbib}

\begin{document}

\title{Microwave Down-Conversion with an Impedance-Matched $\mathbf{\Lambda}$ System in Driven Circuit QED}

\author{K. Inomata} \email[kunihiro.inomata@riken.jp]{}
\affiliation{RIKEN Center for Emergent Matter Science (CEMS), Wako, Saitama 351-0198, Japan}

\author{K. Koshino}
\affiliation{College of Liberal Arts and Sciences, Tokyo Medical and Dental University, Ichikawa, Chiba 272-0827, Japan}

\author{Z. R. Lin}
\affiliation{RIKEN Center for Emergent Matter Science (CEMS), Wako, Saitama 351-0198, Japan}

\author{W. D. Oliver}
\affiliation{MIT Lincoln Laboratory, 244 Wood Street, Lexington, Massachusetts 02420, USA}

\author{J. S. Tsai}
\affiliation{RIKEN Center for Emergent Matter Science (CEMS), Wako, Saitama 351-0198, Japan}
\affiliation{NEC Smart Energy Research Laboratories, Tsukuba, Ibaraki 305-8501, Japan}

\author{Y. Nakamura}
\affiliation{RIKEN Center for Emergent Matter Science (CEMS), Wako, Saitama 351-0198, Japan}
\affiliation{Research Center for Advanced Science and Technology (RCAST),
The University of Tokyo, Meguro-ku, Tokyo 153-8904, Japan}

\author{T. Yamamoto} \email[t-yamamoto@fe.jp.nec.com]{}
\affiliation{RIKEN Center for Emergent Matter Science (CEMS), Wako, Saitama 351-0198, Japan}
\affiliation{NEC Smart Energy Research Laboratories, Tsukuba, Ibaraki 305-8501, Japan}

\date{\today}

\begin{abstract}
By driving a dispersively coupled qubit-resonator system, we realize an ``impedance-matched" $\Lambda$ system that has two identical radiative decay rates from the top level and interacts with a semi-infinite waveguide. 
It has been predicted that a photon input from the waveguide deterministically induces a Raman transition in the system and switches its electronic state. 
We confirm this through microwave response to a continuous probe field, observing near-perfect ($99.7\%$) extinction of the reflection and highly efficient ($74\%$) frequency down-conversion. 
These proof-of-principle results lead to deterministic quantum gates between material qubits and microwave photons and open the possibility for scalable quantum networks interconnected with waveguide photons. 
\end{abstract}

\pacs{03.67.Lx, 85.25.Cp, 42.50.Pq}

\maketitle
In one-dimensional (1D) optical systems, interference between an incident photon field and radiation from a quantum emitter (natural or artificial atom) is drastically enhanced due to the low dimensionality~\cite{Collett84,Shen05_2}. This may be contrasted with the three-dimensional case, where the spatial mode mismatch between the incident and scattered fields prevents perfect interference \cite{Zumofen08}. 
In particular, when the quantum emitter is coupled to the end of a semi-infinite waveguide and when its excited state has two radiative decay paths (i.e., a so-called $\Lambda$ or $\Delta$-type three-level system) with equal decay rates, a resonant incident photon into the emitter deterministically induces a Raman transition, and is never reflected due to destructive interference with the re-emitted photon~\cite{Koshino09}.
This phenomenon is called ``impedance matching", in analogy with the suppression of wave reflection in an electric circuit terminated by its characteristic impedance~\cite{Afzelius10}. 

Artificial atoms in superconducting circuits have proven to be versatile quantum mechanical systems for realizing a variety of intriguing quantum optical phenomena. 
In circuit quantum electrodynamics (QED) \cite{Blais04,Wallraff05}, strong coupling of a superconducting qubit with a resonator photon is readily achieved. Moreover, an artificial atom coupled directly with a microwave transmission line demonstrates near-perfect reflection of the incident  field \cite{Oleg10,Hoi11}.
Recently, we have theoretically shown that an impedance-matched $\Lambda$ system can be implemented by using the dressed states of a driven circuit-QED system~\cite{Koshino13}.
Although $\Lambda$ systems have been implemented with a flux qubit by using the lowest three levels of its asymmetric double-well potential~\cite{Grajcar08,Valenzuela06}, realizing an impedance-matched $\Lambda$ system has remained elusive. 
Here, we experimentally demonstrate impedance matching in the driven circuit-QED system.
Using this system, we demonstrate near-perfect absorption of the incident microwave and its frequency down-conversion with a conversion efficiency of 74$\%$. 
These results and their associated agreement with our model calculations indicate that each incident microwave photon deterministically induces a Raman transition in the $\Lambda$ system and excites the qubit.
Compared to the recently demonstrated high-efficiency capturing of an itinerant microwave pulse \cite{Wenner13}, the present scheme does not require any precise pulse-shaping of the input photons, nor the time-dependent control of system parameters such as the transition energies of the artificial atom or its coupling to the waveguide. 
The down-conversion process is also accompanied by a flip of the qubit state, enabling its applications to quantum logic gates and memories~\cite{Koshino10} and single-photon detectors in the microwave domain~\cite{Chen11}. This discriminates the present scheme from other frequency conversion circuits such as in Ref. \onlinecite{Abdo13}.

We consider a flux qubit coupled to a coplanar waveguide (CPW) resonator (Fig.~1).
In this coupled system, a dispersive frequency shift is enhanced by the effect of the straddling regime~\cite{Koch07} and capacitive coupling~\cite{Inomata12,Yamamoto14}.
Below we treat the qubit as a two-level system since the higher energy levels of the qubit have little effect except for the enhanced dispersive shifts.
The level structure of the coupled system is depicted by the Jaynes-Cummings ladder in Fig.~2(a). 
In the present study, only the lowest four levels are relevant because of the weak probe field applied to the resonator. 
We apply a drive field (frequency $\omega_{\rm d}$ and power $P_{\rm d}$) to the qubit to generate the qubit-resonator dressed states and a probe field (frequency $\omega_{\rm p}$ and power $P_{\rm p}$) to the CPW resonator to observe their microwave response. 
By switching to a frame rotating at $\omega_{\rm d}$, the eigenenergies are given as 
\begin{eqnarray}\label{gEng}
\omega_{\vert g, n\rangle} &=& n(\omega_{\rm r} - \omega_{\rm d}), \\
\omega_{\vert e, n\rangle} &=& \omega_{\rm ge} - \omega_{\rm d} + n(\omega_{\rm r} - \omega_{\rm d}-2\chi),
\end{eqnarray}
where $\omega_{\rm r}$, $\omega_{\rm ge}$, and $\chi$ are the resonant frequency of the CPW resonator, the transition frequency of the qubit from the ground state $\vert g \rangle$ to the first excite state $\vert e \rangle$, and the dispersive frequency shift, respectively. 
Note that $\omega_{\rm r}$ and $\omega_{\rm ge}$ are not their bare frequencies but the renormalized ones including the dispersive shifts~\cite{Supp_Note}.
By choosing $\omega_{\rm d}$ within the range of 
$\omega_{\rm ge}-2\chi<\omega_{\rm d}<\omega_{\rm ge}$, 
the system can be set into the ``nesting regime", where the level structure becomes nested, i.e.,
$\omega_{\vert g, 0\rangle}<\omega_{\vert e, 0\rangle}<\omega_{\vert e, 1\rangle}<\omega_{\vert g, 1\rangle}$~\cite{Koshino13}. 
The qubit drive mixes the lower (higher) two levels in Fig.~2(b) with each other to form dressed states $\vert \widetilde{1} \rangle$ and $\vert \widetilde{2} \rangle$ ($\vert \widetilde{3} \rangle$ and $\vert \widetilde{4} \rangle$). 
Under a proper choice of the drive power, the two radiative decay rates from $\vert \widetilde{3} \rangle$ or $\vert \widetilde{4} \rangle$ become identical (either $\tilde{\kappa}_{31} = \tilde{\kappa}_{32}$ or $\tilde{\kappa}_{41} = \tilde{\kappa}_{42}$). Then the coupled system functions as an impedance-matched $\Lambda$ system, where the ground state $\vert G \rangle=\vert \widetilde{1} \rangle$, the middle state $\vert M \rangle=\vert \widetilde{2} \rangle$, and the excited state $\vert E \rangle=\vert \widetilde{3} \rangle$ or, alternatively, $\vert \widetilde{4} \rangle$ [see Fig.~2(c)]. 
For such a configuration, quantum interference ensures that incident photons resonant with the $\vert G \rangle \rightarrow \vert E \rangle$ transition will deterministically induce a Raman transition of $\vert G \rangle \rightarrow \vert E \rangle \rightarrow \vert M \rangle$. This can be observed in microwave spectroscopy as perfect absorption of the incident field and frequency down-conversion of the reflected field~\cite{Koshino13}.
 \begin{figure}[]
 \includegraphics[width=8.3cm]{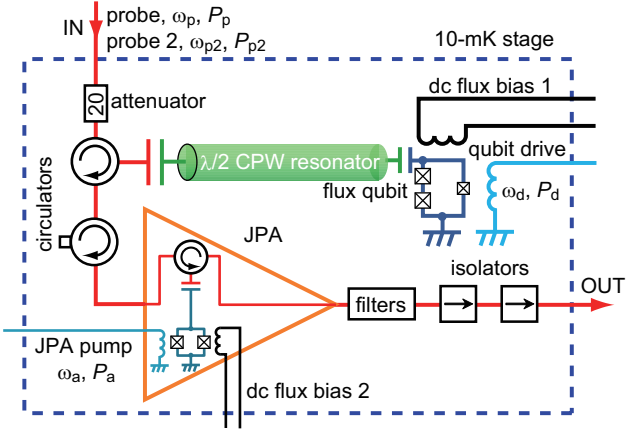}
 \label{circ}
 \caption{(Color online) 
 Experimental setup at the $10$-mK stage of a dilution refrigerator. The resonator+qubit chip and the JPA chip are mounted in separate sample packages with independent coils for dc flux bias.}
 \end{figure}

A schematic of the measurement setup at the 10-mK stage of a dilution refrigerator is shown in Fig.~1. The setup includes two circuits fabricated on separate chips: a flux qubit capacitively coupled to a half-wavelength CPW resonator and a flux-driven Josephson parametric amplifier (JPA). They are separately mounted in microwave-tight packages equipped with an independent coil for the flux bias and connected with each other via three circulators in series~\cite{Lin13}. The qubit chip is the same as the one used in Ref.~\onlinecite{Inomata12}. The qubit is biased with a half flux quantum where $\omega_{\rm ge}/2\pi=$ 5.461~GHz is insensitive to low frequency flux noise to first order.
The resonator frequency $\omega_{\rm r}/2\pi$ is 10.678~GHz when the qubit is in the $\vert g\rangle$ state. It is shifted by $-2\chi/2\pi =-80~{\rm MHz}$ when the qubit is in the $\vert e\rangle$ state.

The JPA consists of a $\lambda /4$ CPW resonator terminated by a SQUID, whose design and fabrication process are reported in Ref.~\onlinecite{Yamamoto08}. 
The JPA is used to amplify the down-converted microwave field from the impedance-matched $\Lambda$ system. Except during this particular measurement, the JPA is kept off. Namely, the amplifier pump field is turned off and the resonant frequency of its resonator is detuned from $\omega_{\rm r}$ so that the JPA acts as a perfect mirror.

We first measure the reflection coefficient $r$ of the qubit-resonator coupled system as a function of $\omega_{\rm p}$ and $P_{\rm d}$ using a vector network analyzer (VNA) [Fig.~3(a)]. Microwave power levels stated in this paper are referred to the corresponding ports on the sample chip~\cite{Supp_Note}. The probe field with $P_{\rm p} = -146.2~{\rm dBm}$, corresponding to an average photon number of 0.013 in the resonator, is generated by the VNA, while the qubit drive field is applied from another microwave source. The qubit is continuously driven at $\omega_{\rm d}$ detuned from $\omega_{\rm ge}$ by 
$\delta\omega_{\rm d} \equiv \omega_{\rm d} - \omega_{\rm ge} = 2\pi \times (-64)~{\rm MHz}$ ($|\delta \omega_{\rm d}|<2\chi$), 
so that the system is in the nesting regime.
 \begin{figure}
 \includegraphics[width=8.5cm]{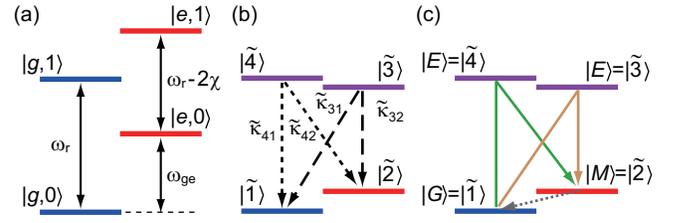}
 \label{elevel}
 \caption{(Color online) Energy-level diagram of the coupled system. 
 (a) Lowest four energy levels of the qubit-resonator coupled system. $\vert k,l \rangle$ denotes the eigenstates of the Jaynes-Cummings Hamiltonian, which are close to the product states $\vert k \rangle_{\rm q} \vert l \rangle_{\rm r}$ of the qubit and the resonator, where $k=g, e$ and $l=0, 1, \cdots$.  
 (b)\nobreak~Dressed-state energy levels in the frame rotating at $\omega_{\rm d}$. $\vert \widetilde{i} \rangle$ represents the dressed state of the qubit-resonator coupled system, and $\tilde{\kappa}_{ij}$ is the radiative decay rate for the $\vert \widetilde{i}\rangle \rightarrow \vert \widetilde{j}\rangle$ transition. The ``nesting regime" is realized when $\omega_{\rm ge}-2\chi < \omega_{\rm d} < \omega_{\rm ge}$. 
 (c) Raman processes in the impedance-matched $\Lambda$ system (solid arrows). The $\vert \widetilde{2} \rangle \rightarrow \vert \widetilde{1} \rangle$ decay (dotted line) is mainly caused by qubit relaxation.}
 \end{figure}

For a weak drive $(P_{\rm d}<-90~{\rm dBm})$, the probe field is fully reflected. In contrast, as we increase $P_{\rm d}$ to $P_{\rm d4} \equiv -84$~dBm or $P_{\rm d3} \equiv -77$~dBm, the reflected probe field vanishes at certain frequencies. Figure~3(b) shows a cross-section of Fig.~3(a) at $P_{\rm d4}$, presenting a dip with $-25$~dB ($99.7\%$) suppression at $\omega_{\rm p}/2\pi=10.681$~GHz ($\approx \omega_{\rm r}/2\pi$). For comparison, we theoretically calculate $r$ and the radiative decay rates $\tilde{\kappa}_{ij}$ for the $\vert \widetilde{i} \rangle \rightarrow \vert \widetilde{j} \rangle$ transition [Figs.~3(c) and (d)]~\cite{Koshino13}.
Here, we use a decay rate of a resonator photon $\kappa=\kappa_1+\kappa_2=2\pi \times 16.4$~MHz, where $\kappa_1=0.95\kappa$ and $\kappa_2=0.05\kappa$ are radiative and non-radiative components, respectively, and the qubit energy decay rate $T_1^{-1} = 2\pi \times 0.227$~MHz, all of which are determined from independent measurements~\cite{Supp_Note}.
The radiative decay rate of the qubit into the drive port is set to $\gamma_{\rm c}/2\pi=0.6$~kHz to adjust the horizontal scale in Fig.~3(c).
We find fairly good agreement between Figs.~3(a) and (c). In Fig.~3(c), we draw the transition frequencies $\tilde{\omega}_{ij}$ between the state $\vert \widetilde{j} \rangle$ and $\vert \widetilde{i} \rangle$ by dashed curves. They indicate that the probe fields are efficiently absorbed at $(P_{\rm d}, \omega_{\rm p}) = (P_{\rm d4}, \tilde{\omega}_{41})$ and $(P_{\rm d3}, \tilde{\omega}_{31})$.
 \begin{figure}
 \includegraphics[width=8.5cm]{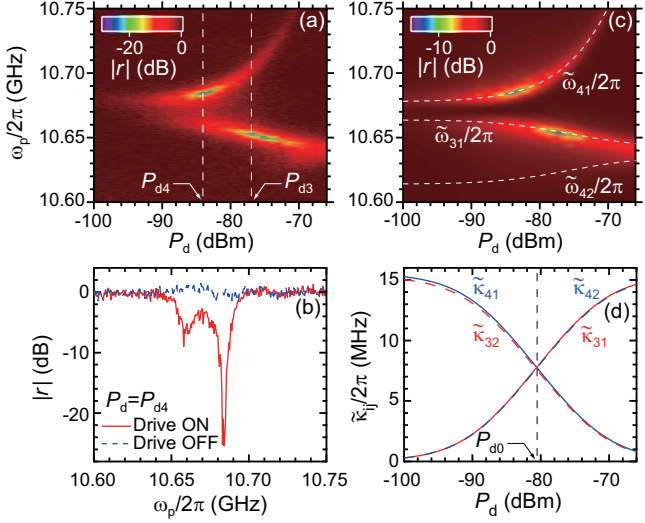}
 \label{SRF}
 \caption{(Color online) Perfect absorption of the probe field. 
 (a) Reflection coefficient $|r|$ as a function of the probe frequency $\omega_{\rm p}$ and the qubit drive power $P_{\rm d}$. The probe power is fixed at $-146.2$~dBm, and the detuning $\delta\omega_{\rm d}$ ($\equiv \omega_{\rm d} - \omega_{\rm ge}$) is fixed at $2\pi \times (-64)$~MHz. The dashed lines show the drive powers ($P_{\rm d3}$ and $P_{\rm d4}$) where the depth of the dips is maximized.
 (b) $|r|$ as a function of $\omega_{\rm p}$ for $P_{\rm d}=P_{\rm d4}$. The red solid and blue dashed lines depict the data with the qubit drive on and off, respectively. 
 (c)\nobreak~Numerical simulations corresponding to (a). The dashed lines indicate transition frequencies between the dressed states. 
 (d)\nobreak~Calculated radiative decay rates $\tilde{\kappa}_{ij}$ as a function of $P_{\rm d}$. The dashed line shows the drive power $P_{\rm d0}$, where $\tilde{\kappa}_{41} = \tilde{\kappa}_{42}$ and $\tilde{\kappa}_{31} = \tilde{\kappa}_{32}$.}
 \end{figure}

Figure~3(d) shows the calculated radiative decay rates from the state $\vert \widetilde{4}\rangle$ ($\tilde{\kappa}_{41}$ and $\tilde{\kappa}_{42}$) and $\vert \widetilde{3}\rangle$ ($\tilde{\kappa}_{31}$ and $\tilde{\kappa}_{32}$) as a function of $P_{\rm d}$. The two radiative decay rates from each state ideally become identical at $P_{\rm d} = P_{\rm d0} \equiv -80.6$~dBm.
In theory, the impedance matching is expected to occur at $(P_{\rm d}, \omega_{\rm p}) = (P_{\rm d0}, \tilde{\omega}_{41})$ and $(P_{\rm d0}, \tilde{\omega}_{31})$, where $\tilde{\kappa}_{41} = \tilde{\kappa}_{42}$ and $\tilde{\kappa}_{31} = \tilde{\kappa}_{32}$~\cite{Koshino13}. 
In the actual system, however, inadvertent population of the $\vert M \rangle$ state due to the continuous probe field and the intrinsic loss of the resonator ($\kappa_2$) weaken the radiation from the $\Lambda$ system, resulting in an imperfect destructive interference with the reflected wave. To compensate this effect, and thereby recover complete cancellation,  
the elastic component of the radiation (the $\vert E \rangle \rightarrow \vert G \rangle$ decay) should be slightly larger than the inelastic component (the $\vert E \rangle \rightarrow \vert M \rangle$ decay).
As a consequence, the impedance matching occurs in practice at $(P_{\rm d}, \omega_{\rm p}) = (P_{\rm d4}, \tilde{\omega}_{41})$ and $(P_{\rm d3}, \tilde{\omega}_{31})$, where $\tilde{\kappa}_{41} > \tilde{\kappa}_{42}$ and $\tilde{\kappa}_{31} > \tilde{\kappa}_{32}$, respectively, as demonstrated by the pronounced dips in Figs.~3(a) and (c). The impedance-matched $\Lambda$ systems with $\vert E\rangle = \vert \widetilde{4} \rangle$ and $\vert E\rangle = \vert \widetilde{3} \rangle$ [see Fig.~2(c)] are realized, correspondingly. 

Next, we demonstrate that we can control the energy levels of the impedance-matched $\Lambda$ system by changing $\delta\omega_{\rm d}$.
For this purpose, we first conduct measurements similar to Fig.~3(a) for each $\delta\omega_{\rm d}$, and determine $P_{\rm d}$ for the impedance matching condition under the continuous probe field (probe~1) with a fixed $P_{\rm p}$ of $-141.2$~dBm and $\omega_{\rm p}$ of $2\pi \times10.681$~GHz ($=\tilde{\omega}_{41}$). Then, we probe the $\Lambda$ system with $\vert E \rangle = \vert \widetilde{4} \rangle$ by using the VNA (probe~2) with a power $P_{\rm p2}=-135~{\rm dBm}$ and frequency $\omega_{\rm p2}$.
 \begin{figure}
 \includegraphics[width=7.2cm]{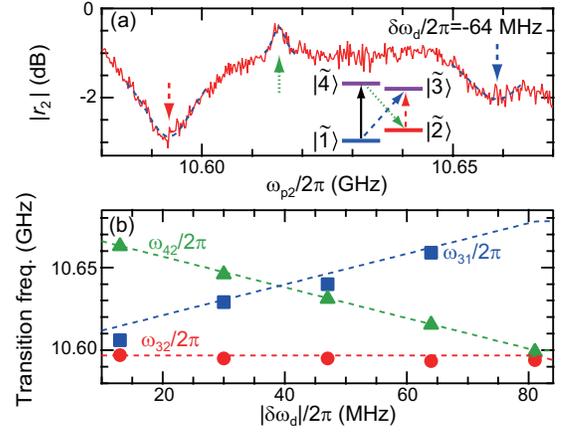}
 \label{SCW}
 \caption{(Color online) Tunability of the $\Lambda$ system.
 (a)\nobreak~Reflection coefficient $|r_2|$ of probe~2 as a function of $\omega_{\rm p2}$. Here, $\omega_{\rm p}/2\pi=10.681$~GHz, $P_{\rm p}=-141.2$~dBm, $P_{\rm p2}=-135$~dBm, and $\delta\omega_{\rm d}/2\pi=-64$~MHz. The dashed curves are fits with Lorentzian functions and the arrows represent center frequencies of the peak and dips. They indicate the transitions depicted in the energy-level diagram in the inset by corresponding arrows. Vertical black arrow in the inset represents the transition induced by probe~1.
 (b) Center frequencies of the observed peaks and dips as a function of $\delta\omega_{\rm d}$. The solid symbols and dashed curves represent the measured and calculated transition frequencies, respectively. 
 }
 \end{figure}

Figure~4(a) shows $|r_2|$ of probe~2 as a function of $\omega_{\rm p2}$ for $\delta\omega_{\rm d}/2\pi=-64$~MHz. 
As indicated by arrows in Fig.~4(a), a peak and two dips are observed.
The origin of these signals can be qualitatively understood by the energy-level diagram shown in the inset of Fig.~4(a). 
Since probe~1 continuously drives the $\vert \widetilde{1} \rangle \rightarrow \vert \widetilde{4} \rangle$ transition and populates the $\vert \widetilde{4} \rangle$ state, probe~2 stimulates the $\vert \widetilde{4} \rangle \rightarrow \vert \widetilde{2} \rangle$ transition when it is tuned to $\tilde{\omega}_{\rm 42}$.
This is observed as the peak at $\omega_{\rm p2}/2\pi=10.615$~GHz, corresponding to the frequency down-conversion of probe~1.
The dips at $\omega_{\rm p2}/2\pi=$10.659~GHz and 10.593~GHz result from absorption of probe~2 when it is tuned to $\tilde{\omega}_{31}$ and $\tilde{\omega}_{32}$, respectively.
In Fig.~4(b), we plot center frequencies of these peaks and dips for different $\delta\omega_{\rm d}$'s. 
As expected, $\tilde{\omega}_{31}$ ($\tilde{\omega}_{42}$) is an increasing (a decreasing) function of $|\delta\omega_{\rm d}|$, whereas $\tilde{\omega}_{32}$ is independent of $\delta\omega_{\rm d}$. To understand this result quantitatively, we calculated the transition frequencies. They are plotted by dashed curves in Fig.~4(b), and agree well with the experimental data.
Note that the impedance matching is not realized any more at $|\delta \omega_{\rm d}|/2\pi = 81$~MHz in Fig.~4(b) because $|\delta \omega_{\rm d}|>2\chi$.

In the above measurement, we indirectly observed the frequency down-conversion of probe~1 as a stimulated emission peak in $|r_2|$. Now we directly measure the down-converted signal from the impedance-matched $\Lambda$ system with $\vert E \rangle = \vert \widetilde{4} \rangle$. In this measurement, we use a flux-driven JPA to amplify the weak signal. The JPA is operated in a nondegenerate mode at 10.6145~GHz, with a signal gain of 21~dB and a bandwidth of 2$\pi \times 3.3$~MHz. The qubit is driven at $\delta \omega_{\rm d}/2\pi = -64$~MHz, and a probe power $P_{\rm p}$ of $-146.2$~dBm is injected. To improve the signal-to-noise ratio, which is below unity even with the JPA, we repeatedly turn on and off the qubit drive~\cite{JPA_Note}, and average the difference in the spectral density $4\times 10^4$ times using a spectrum analyzer.
 \begin{figure}
 \includegraphics[width=8.5cm]{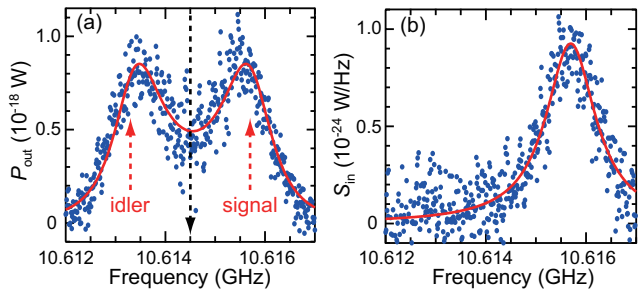}
 \label{DCS}
 \caption{(Color online) Down-conversion of the microwave field. (a) Output power of the JPA. Blue dots are the measurement data and the solid line is a fit with Eq.(\ref{Spec}). A JPA with $G_{\rm s}=21$~dB and a bandwidth of $2\pi \times 3.3$~MHz was used. The black arrow indicates the band center of the JPA.
 (b) Extracted power spectral density of the down-converted signal. The solid line shows calculated $S_{\rm in}$ with parameters obtained by the fitting in (a).} 
 \end{figure}

Figure~5(a) shows the obtained down-converted spectrum. Two peaks appear symmetrically with respect to the band center of the JPA, corresponding to the signal and idler components of the JPA output.
The JPA output power is given by
\begin{equation}\label{Spec}
P_{\rm out}(\omega)=[G_{\rm s}(\omega)S_{\rm in}(\omega) + G_{\rm i}(\omega_{\rm a}-\omega)S_{\rm in}(\omega_{\rm a}-\omega)]B,
\end{equation}
where $\omega_{\rm a}$ is the JPA pump frequency, $B=2\pi \times 10$~kHz is the resolution bandwidth of the spectrum analyzer, and $S_{\rm in}$ is the power spectral density of the down-converted signal, which we assume to have a Lorentzian lineshape with a center frequency of $\omega_{\rm s}$. $G_{\rm s}$ and $G_{\rm i}$ are the signal and idler gains of the JPA which are almost identical within the measurement range. By fitting the measured spectrum with Eq.~(\ref{Spec}) [solid curve in Fig.~5(a)], we extracted $S_{\rm in}(\omega)$ as shown in Fig.~5(b), and obtained $\omega_{\rm s}=2\pi \times 10.6157~$GHz and $\delta \omega=2\pi \times 1.210$~MHz, where $\delta \omega$ is a linewidth of the signal. The signal frequency $\omega_{\rm s}$ agrees very well with the frequency of the stimulated emission peak in Fig.~4(a),
while $\delta \omega$ is six times larger than the expected value (the $\vert \widetilde{2}\rangle \rightarrow \vert \widetilde{1}\rangle$ decay rate, which roughly coincides with $T_1^{-1}$)~\cite{Koshino13_2}. 
A possible reason for this is inhomogeneous broadening due to fluctuations of $\omega_{\rm ge}$ during the measurement which takes nearly three hours.

From the data in Fig.~5(b), we estimate the down-conversion efficiency $\eta$, which is defined by the flux of down-converted photons normalized by the input flux. 
The signal power $P_{\rm s}$ obtained by integrating the down-converted signal in Fig.~5(b) is $(1.77 \pm 0.20) \times 10^{-18}$~W, while the input probe power $P_{\rm p}$ is $10^{-17.62}$~W. From these values, $\eta~(=P_{\rm s}\omega_{\rm p}/P_{\rm p}\omega_{\rm s})$ is determined to be $74 \pm 8\%$, where the uncertainty comes from the inaccuracy in the estimation of the total gain ($\pm 0.5$~dB) in the output microwave lines. On the other hand, a theoretical estimation based on the experimental parameters gives $\eta$ of 68$\%$~\cite{Supp_Note}.
The loss of $\eta$ is attributed to the intrinsic loss of the resonator and the incomplete initialization of the ground state $\vert \widetilde{1} \rangle$ due to continuous excitation by the probe.
Although it is difficult to confirm in the present setup due to insufficient signal-to-noise ratio, we expect for a weaker continuous wave or single microwave photons as the probe, nearly complete down-conversion of $\eta \simeq 95\%$ ($=\kappa_1 / \kappa$).

In conclusion, we experimentally realized an impedance-matched $\Lambda$ system using dressed-state engineering of a driven circuit-QED system, here a superconducting flux qubit and CPW resonator connected to a semi-infinite transmission line.
The results lead to deterministic quantum gates between material qubits and microwave photons and open the possibility for scalable quantum networks interconnected with waveguide photons.

\vspace{0.5 cm}
We would like to thank O. Astafiev for technical help, V. Bolkhovsky and G. Fitch for assistance with the JPA fabrication at MIT-LL.
Sputtered Nb films were fabricated in the clean room for analogue-digital superconductivity (CRAVITY) in the National Institute of Advanced Industrial Science and Technology (AIST).
This work was supported by the Funding Program for World-Leading Innovative R$\&$D on Science and Technology (FIRST), the Grant-in-Aid for Scientific Research Program for Quantum Cybernetics of the Ministry of Education, Culture, Sports, Science, and Technology (MEXT), Japan, and the NICT Commissioned Research.  

\newpage

\appendix

\newcommand{\beq}{\begin{equation}}
\newcommand{\eeq}{\end{equation}}
\newcommand{\bea}{\begin{eqnarray}}
\newcommand{\eea}{\end{eqnarray}}
\newcommand{\bec}{\begin{center}}
\newcommand{\enc}{\end{center}}
\newcommand{\bfr}{\begin{flushright}}
\newcommand{\efr}{\end{flushright}}
\newcommand{\alp}{\alpha}
\newcommand{\bet}{\beta}
\newcommand{\om}{\omega}
\newcommand{\tom}{\widetilde{\omega}}
\newcommand{\tkap}{\widetilde{\kappa}}
\newcommand{\tgam}{\widetilde{\gamma}}
\newcommand{\tsig}{\widetilde{\sigma}}
\newcommand{\tg}{\widetilde{g}}
\newcommand{\ta}{\widetilde{a}}
\newcommand{\tb}{\widetilde{b}}
\newcommand{\tc}{\widetilde{c}}
\newcommand{\td}{\widetilde{d}}
\newcommand{\ti}{\widetilde{i}}
\newcommand{\tj}{\widetilde{j}}
\newcommand{\tone}{\widetilde{1}}
\newcommand{\ttwo}{\widetilde{2}}
\newcommand{\tthree}{\widetilde{3}}
\newcommand{\tfour}{\widetilde{4}}
\newcommand{\Om}{\Omega}
\newcommand{\Lam}{\Lambda}
\newcommand{\eps}{\epsilon}
\newcommand{\ve}{\varepsilon}
\newcommand{\kap}{\kappa}
\newcommand{\gam}{\gamma}
\newcommand{\G}{\Gamma}
\newcommand{\s}{\sigma}
\newcommand{\da}{a^{\dagger}}
\newcommand{\ds}{s^{\dagger}}
\newcommand{\dR}{R^{\dagger}}
\newcommand{\la}{\langle}
\newcommand{\ra}{\rangle}
\newcommand{\rmi}{{\rm i}} 
\newcommand{\cH}{{\cal H}}
\newcommand{\cN}{{\cal N}}
\newcommand{\cV}{{\cal V}}
\newcommand{\cG}{{\cal G}}
\newcommand{\tred}{\textcolor{red}}
\newcommand{\tblue}{\textcolor{blue}}
\newcommand{\tgrn}{\textcolor{green}}
\newcommand{\tcyan}{\textcolor{cyan}}
\newcommand{\tmag}{\textcolor{magenta}}

\section{Microwave response theory of impedance-matched $\Lambda$ system}

\subsection{Dressed states of a driven circuit QED system}

\begin{figure}[h]
\includegraphics[width=7.4cm]{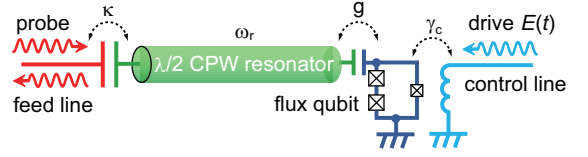}
\caption{Circuit diagram of the qubit-resonator coupled system.}
\label{fig:S1}
\end{figure}

Figure~\ref{fig:S1} is the circuit diagram of the device considered.
It consists of a semi-infinite waveguide (red),
a resonator (green), a superconducting flux qubit (blue), and 
a control line to apply a qubit drive (light blue).
The flux qubit is generally used as a two-level system ($|{g}\ra$ and $|{e}\ra$).
However, we here include the second excited state $|{f}\ra$ in the model
to correctly evaluate the dispersive level shifts.
Putting $\hbar=v=1$, where $v$ is the microwave velocity in the waveguide,
the Hamiltonian of the qubit-resonator system is given by
\bea
\cH_{\rm sys}(t) &=& \cH_{\rm JC}+\cH_{\rm dr}(t),
\\
\nonumber
\cH_{\rm JC} &=& 
\bar{\om}_{\rm ge}\s_{\rm ee}+\bar{\om}_{\rm gf}\s_{\rm ff}+\bar{\om}_{\rm r}a^{\dag}a\\
&&+g_{\rm ge}(a^{\dag}\s_{\rm ge}+\s_{\rm eg}a)
+g_{\rm ef}(a^{\dag}\s_{\rm ef}+\s_{\rm fe}a), \hspace{5 mm}
\\
\cH_{\rm dr}(t) &=& \sqrt{\gam_{\rm c}}\left[E(t)\s_{\rm eg}+E^*(t)\s_{\rm ge}\right],
\eea
where $\cH_{\rm JC}$ is the Jaynes-Cummings Hamiltonian for the qubit and the resonator
and $\cH_{\rm dr}(t)$ describes the qubit drive.
The operators are defined as follows:
$\s_{\rm pq}=|{\rm p}\ra\la{\rm q}|$ is the qubit transition operator
and $a$ is the annihilation operator of the resonator photon.
The meanings of the parameters are as follows:
$\bar{\om}_{\rm ge}$ and $\bar{\om}_{\rm gf}$ are the {\it bare} eigenfrequencies 
of $|{e}\ra$ and $|{f}\ra$ measured from $|{g}\ra$,
$\bar{\om}_{\rm r}$ is the {\it bare} resonance frequency of the resonator, 
$g_{\rm ge}$ and $g_{\rm ef}$ are the coupling constants between the qubit and the resonator,
and $\gam_{\rm c}$ is the coupling constant between the qubit and the drive field.
The drive field is monochromatic and tuned close to $\om_{\rm ge}$, namely,
$E(t)=E_{\rm d}e^{-{\rm i}\om_{\rm d}t}$ and $\om_{\rm d}\simeq \bar{\om}_{\rm ge}$. 
We neglect 
the $|{e}\ra\leftrightarrow|{f}\ra$ drive term, 
since the drive frequency is detuned largely from $\bar{\om}_{\rm ef}$.

The present qubit-resonator system is in the dispersive regime,
where the qubit-resonator detunings dominate their couplings,
i.e., $|\bar{\om}_{\rm ge}-\bar{\om}_{\rm r}| \gg g_{\rm ge}$ and 
$|\bar{\om}_{\rm ef}-\bar{\om}_{\rm r}| \gg g_{\rm ef}$.
In this regime, the couplings do not significantly mix the quantum states
but renormalize their frequencies. 
After renormalization, the lowest four levels of the qubit-resonator system 
($|g,0\ra$, $|g,1\ra$, $|e,0\ra$ and $|e,1\ra$)
are relevant in this study. 
Their renormalized frequencies are given by
$\om_{|{\rm g},n\ra}=n\om_{\rm r}$ and 
$\om_{|{\rm e},n\ra}=\om_{\rm ge}+n(\om_{\rm r}-2\chi)$
[Fig.~2(a) of the main text], where
\bea
\om_{\rm ge} &=& 
{\textstyle
\bar{\om}_{\rm ge}-\frac{g_{\rm ge}^2}{\bar{\om}_{\rm r}-\bar{\om}_{\rm ge}},
}
\\
\om_{\rm r} &=& 
{\textstyle
\bar{\om}_{\rm r}+\frac{g_{\rm ge}^2}{\bar{\om}_{\rm r}-\bar{\om}_{\rm ge}},
}
\\
\chi &=& 
{\textstyle
\frac{g_{\rm ge}^2}{\bar{\om}_{\rm r}-\bar{\om}_{\rm ge}}+
\frac{g_{\rm ef}^2}{2(\bar{\om}_{\rm ef}-\bar{\om}_{\rm r})}.
}
\eea
$\chi$ is called the dispersive frequency shift. 
We assumed the following parameter values:
$\bar{\om}_{\rm ge}/2\pi=5.468$~GHz,
$\bar{\om}_{\rm gf}/2\pi=19.362$~GHz,
$\bar{\om}_{\rm r}/2\pi=10.671$~GHz,
$g_{\rm ge}/2\pi=0.197$~GHz, and
$g_{\rm ef}/2\pi=0.458$~GHz.
This reproduces the experimentally observed parameters:
$\om_{\rm ge}/2\pi=5.461$~GHz,
$\om_{\rm r}/2\pi=10.678$~GHz, and
$2\chi/2\pi=80$~MHz.
We note that $\omega_{\rm ge}$ and $\omega_{\rm r}$ have been slightly varied 
from the previous values described in Ref.~\onlinecite{Inomata12} after thermal cycling.
The coupling between the qubit and the drive port 
is set as $\gamma_{\rm c}/2\pi=0.6$~kHz 
to adjust the horizontal scale in Fig.~3(c) in the main text.

In the frame rotating at $\om_{\rm d}$, the system Hamiltonian becomes static.
It is given by~\cite{Koshino13}
\bea
\cH_{\rm sys} &=& \cH_{\rm JC}+\cH_{\rm dr},
\\
\cH_{\rm JC} &=& 
\sum_{n=0,1}
\left( \om_{|{\rm g},n\ra}|g,n\ra\la g,n|
+\om_{|{\rm e},n\ra}|e,n\ra\la e,n| \right),\hspace{5 mm}
\\
\cH_{\rm dr} &=& 
\sum_{n=0,1} 
\sqrt{\gam_{\rm c}}E_{\rm d}
\left(|g,n\ra\la e,n| +
|e,n\ra\la g,n|\right),
\eea
where 
\bea
\om_{|{\rm g},n\ra} &=& n(\om_{\rm r}-\om_{\rm d}),
\label{eq:omgn}
\\
\om_{|{\rm e},n\ra} &=& \om_{\rm ge}-\om_{\rm d}+n(\om_{\rm r}-\om_{\rm d}-2\chi).
\label{eq:omen}
\eea
From Eqs.~(\ref{eq:omgn}) and (\ref{eq:omen}), we readily observe that 
the level structure of the four states is nested 
when the drive frequency is chosen to satisfy 
$\om_{\rm ge}-2\chi<\om_{\rm d}<\om_{\rm ge}$ (nesting regime)
and is not nested otherwise (un-nesting regime).

We refer to the eigenstates of $\cH_{\rm sys}$ as the dressed states.
We label them from the lowest ($j=1,\cdots,4$)
and denote them as $|\widetilde{j}\ra$ and their energies as $\tom_j$
[Fig.~2(b) of the main text]. 
The drive field mixes 
the bare states $|g,0\ra$ and $|e,0\ra$
($|g,1\ra$ and $|e,1\ra$)
to form the dressed states $|\tone\ra$ and $|\ttwo\ra$ 
($|\tthree\ra$ and $|\tfour\ra$).
In Fig.~3(c) of the main text,
the transition frequencies between the dressed states,
$\tom_{ij}=\tom_i-\tom_j$, are drawn as the functions of the drive power, 
$P_{\rm d}=\hbar\om_{\rm d}E_{\rm d}^2$.
Note that the drive frequency $\om_{\rm d}$,
which is subtracted in the rotating frame, is presented in this figure.

\subsection{Radiative and nonradiative decay rates}
The present qubit-resonator system has three damping channels:
(i)~radiative decay from the resonator to the waveguide, 
(ii)~intrinsic loss of the resonator, 
and (iii)~intrinsic loss of the qubit. 
We denote the decay rates associated with these processes by
$\kap_1$, $\kap_2$ and $\gam$, respectively,
and assume that
$\kap_1/2\pi=0.95\times 16.4$~MHz,
$\kap_2/2\pi=0.05\times 16.4$~MHz, and
$\gam/2\pi=0.227$~MHz.
We consider only the $|e\ra \to |g\ra$ decay 
in (iii), 
since the higher qubit levels are mostly unoccupied in this study. 
Furthermore, we consider the qubit radiative decay 
to the drive port to be included in (iii).

In the following part of this subsection, 
we neglect channel~(ii) and denote $\kap_1$ by $\kap$ for simplicity.
The Hamiltonian that describes damping is written,
in the bare-state basis, as
\begin{widetext}
\bea
\cH_{\rm damp} &=& 
\int dk \left[ kb_k^{\dag}b_k + \sqrt{\kap/2\pi}
(a^{\dag}b_k+b_k^{\dag}a) \right]
+
\int dk \left[ kc_k^{\dag}c_k + \sqrt{\gam/2\pi}
(\s_{\rm eg}c_k+c_k^{\dag}\s_{\rm ge}) \right],
\eea
\end{widetext}
where $b_k$ ($c_k$) denotes the waveguide (environment) 
mode with frequency $vk$.
Switching to the dressed-state basis, $\cH_{\rm damp}$ is rewritten as
\begin{widetext}
\bea
\cH_{\rm damp} &=& 
\int dk \left[ kb_k^{\dag}b_k + \sum_{i,j}\sqrt{\tkap_{ij}/2\pi}
(\tsig_{ij}b_k+b_k^{\dag}\tsig_{ji}) \right]
+
\int dk \left[ kc_k^{\dag}c_k + \sum_{i,j}\sqrt{\tgam_{ij}/2\pi}
(\tsig_{ij}c_k+c_k^{\dag}\tsig_{ji}) \right],
\label{eq:Hdamp}
\eea
\end{widetext}
where $\tsig_{ij}=|\ti\ra\la\tj|$.
$\tkap_{ij}$ and $\tgam_{ij}$ respectively denote the radiative 
and nonradiative decay rates for $|\ti\ra \to |\tj\ra$ transition.
They are given by
\bea
\tkap_{ij} &=& \kap|\la\ti|a^{\dag}|\tj\ra|^2,
\\
\tgam_{ij} &=& \gam|\la\ti|\s_{\rm eg}|\tj\ra|^2.
\eea
The radiative decay occurs
from the upper dressed states ($|\tthree\ra$, $|\tfour\ra$)
to the lower states ($|\tone\ra$, $|\ttwo\ra$).
In Fig.~3(d) of the main text, 
the radiative decay rates are plotted as functions of the drive power $P_{\rm d}$
for the case of nesting regime.
The radiative decay rate is divided in two directions,
satisfying the sum rules of
$\tkap_{31}+\tkap_{32}=\kap$ and $\tkap_{41}+\tkap_{42}=\kap$.
On the other hand, the nonradiative decay occurs
within the upper dressed states ($|\tthree\ra$, $|\tfour\ra$)
and the lower dressed states ($|\tone\ra$, $|\ttwo\ra$).
In particular, the $|\ttwo\ra \to |\tone\ra$ relaxation 
has a crucial role in this system, 
since it is the bottleneck process in the down-conversion cycle.
In our experiment, we choose the drive frequency $\om_{\rm d}$ 
close to the lower edge of the nesting regime 
($\om_{\rm d} \simeq \om_{\rm ge}-2\chi$),
in order that the lower two states remain minimally mixed by the drive,
i.e., $|\tone\ra \simeq |g,0\ra$ and $|\ttwo\ra \simeq |e,0\ra$. 
Then, $\tgam_{21}\simeq\gam$ and the others are negligible.

\subsection{Microwave response}
\subsubsection{Hamiltonian and initial state vector}
Here we analyze the microwave response of the qubit-resonator system.
The Hamiltonian of the overall system,
including the waveguide modes and the environment, is given by~\cite{Koshino13_2}
\bea
\cH_{\rm tot} &=& \cH_{\rm sys}+\cH_{\rm damp},
\\
\cH_{\rm sys} &=& \sum_j \tom_j \tsig_{jj},
\eea
where $\cH_{\rm damp}$ is given by Eq.~(\ref{eq:Hdamp}).
The real-space representation of the waveguide field operator is defined by
$\tb_r = (2\pi)^{-1/2}\int dk e^{{\rm i}kr}b_k$.
In this representation, 
the waveguide field interacts with the qubit-resonator system at $r=0$,
and the $r<0$ ($r>0$) region corresponds to the incoming (outgoing) field.
The input and output field operators are defined by
$b_{\rm in}(t)=\tb_{-0}(t)$ and $b_{\rm out}(t)=\tb_{+0}(t)$, respectively.
$c_{\rm in}(t)$ and $c_{\rm out}(t)$ are defined similarly.

As a probe field, we apply a monochromatic classical field 
$E_{\rm p}(r,t)=E_{\rm p}e^{{\rm i}(\om_{\rm p}-\om_{\rm d})(r-t)}$ 
from the waveguide. 
We measure the probe frequency relatively to the drive frequency,
because we work in the rotating frame.
We assume that the system is in its ground state initially (at $t=0$).
Then, the initial state vector of the overall system is written as
\beq
|\Psi_{\rm i}\ra = \cN \exp\left(\int dr E_{\rm p}(r,0) \tb_r^{\dag}\right)|\tone\ra,
\eeq
where $\cN$ is a normalization constant.
Note that 
$b_{\rm in}(t)|\Psi_{\rm i}\ra=E_{\rm p}(0,t)|\Psi_{\rm i}\ra$ 
and $c_{\rm in}(t)|\Psi_{\rm i}\ra=0$.

\subsubsection{Heisenberg equation and input-output relation}
From $\cH_{\rm tot}$, we can rigorously derive the following operator equations.
The input-output relation,
which connect the incoming and outgoing field operators,
is given by
\bea
b_{\rm out}(t) &=& b_{\rm in}(t)-{\rm i}\sum_{m,n}\sqrt{\tkap_{mn}}\tsig_{nm}(t),
\label{eq:ioa}
\eea 
and the Heisenberg equation for the dressed-state transition operator is given by
\begin{widetext}
\beq
\frac{\rm d}{{\rm d}t}\tsig_{ij} =
\sum_{m,n}\left[
\eta^{(1)}_{ijmn}\tsig_{mn}
-{\rm i}\eta^{(2)}_{ijmn}\tsig_{mn}b_{\rm in}(t)
-{\rm i}\eta'^{(2)}_{ijmn}\tsig_{mn}c_{\rm in}(t)
+{\rm i}\eta^{(2)}_{jinm}b^{\dag}_{\rm in}(t)\tsig_{mn}
+{\rm i}\eta'^{(2)}_{jinm}c^{\dag}_{\rm in}(t)\tsig_{mn}
\right],
\label{eq:dsdt}
\eeq
\end{widetext}
where $\eta^{(1)}_{ijmn}={\rm i}(\tom_i-\tom_j)\delta_{im}\delta_{jn}
+\xi^{(1)}_{ijmn}+\xi'^{(1)}_{ijmn}$,
$\eta^{(2)}_{ijmn}=\delta_{im}\sqrt{\tkap_{jn}}-\delta_{jn}\sqrt{\tkap_{mi}}$
and 
$\xi^{(1)}_{ijmn}=\sqrt{\tkap_{mi}\tkap_{nj}}
-\delta_{im}\sum_{\nu}\sqrt{\tkap_{j\nu}\tkap_{n\nu}}/2
-\delta_{jn}\sum_{\nu}\sqrt{\tkap_{i\nu}\tkap_{m\nu}}/2$.
$\eta'^{(2)}_{ijmn}$ and $\xi'^{(1)}_{ijmn}$ are 
obtained by replacing $\tkap$ with $\tgam$.

\subsubsection{Reflection coefficient}
For calculation of the amplitude of reflected field, 
we need the one-point correlation functions of the dressed-state transition operator,
$\la\tsig_{mn}(t)\ra=\la\Psi_{\rm i}|\tsig_{mn}(t)|\Psi_{\rm i}\ra$.
Since we apply a monochromatic probe field,
$\la\tsig_{mn}(t)\ra$ evolves in time as 
$\la\tsig_{mn}(t)\ra=\sum_{q}s_{mn}^{q}e^{{\rm i}q(\om_{\rm p}-\om_{\rm d})t}$ 
in the stationary state, where $q=0,\pm1,\cdots$.
From Eq.~(\ref{eq:dsdt}), we have
\begin{widetext}
\beq
\rmi q (\om_{\rm p}-\om_{\rm d})s_{ij}^{q} = \sum_{m,n}\left[
 \eta^{(1)}_{ijmn} s_{mn}^q
-\rmi E_{\rm p}\eta^{(2)}_{ijmn} s_{mn}^{q+1}
+\rmi E_{\rm p}^*\eta^{(2)}_{jinm} s_{mn}^{q-1}
\right].
\label{eq:sm1}
\eeq
\end{widetext}
The diagonal components of these equations 
are not linearly independent since $\sum_m \s_{mm}=\hat{1}$.
Therefore, we should replace one of them with 
$\sum_m s_{mm}^{q}=\delta_{q0}$.
By solving these simultaneous equations numerically, we determine $s_{mn}^{q}$.
Reliable numerical results are obtained by setting $q=0,\cdots,\pm 3$.

The amplitude of the reflected field at $r=+0$ is determined by 
$E_{\rm out}(t)=\la b_{\rm out}(t) \ra$.
From the input-output relation of Eq.~(\ref{eq:ioa}), we have
\beq
E_{\rm out}(0,t)=E_{\rm p}(0,t)-\rmi\sum_{m,n}\sqrt{\tkap_{mn}}\la\tsig_{nm}(t)\ra.
\eeq
In the stationary state, $E_{\rm out}(t)$ evolves as
$E_{\rm out}(t)=\sum_{q}E_{\rm out}^{q}e^{\rmi q(\om_{\rm p}-\om_{\rm d})t}$. 
The reflection coefficient is defined by $r=E_{\rm out}^{-1}/E_{\rm p}$.
Therefore, $r=1-\rmi\sum_{m,n}\sqrt{\tkap_{mn}}s_{nm}^{-1}/E_{\rm p}$.
The reflection coefficient thus calculated is shown in Fig.~3(c) of the main text.



\subsubsection{Power spectrum}
\label{ssec:ps}
The power spectrum density $S(\om)$ of the reflected field is calculated by
$S(\om)=\hbar\om\times{\rm Re}\int_0^{\infty}d\tau
e^{\rmi(\om-\om_{\rm d})\tau}\la b_{\rm out}^{\dag}(t)b_{\rm out}(t+\tau)\ra/\pi$.
This consists of the coherent and incoherent components.
Our concern lies in the incoherent component,
since the coherent component vanishes in the reflection field 
of an impedance-matched $\Lambda$ system.
The incoherent component is determined by 
the two-point correlation functions of the system operators,
$\la\tsig_{uv}(t),\tsig_{ij}(t+\tau)\ra$,
where $\la A,B \ra=\la AB \ra-\la A\ra\la B\ra$.
Similarly to the one-point correlation function, this quantity is also written as
$\la\tsig_{uv}(t),\tsig_{ij}(t+\tau)\ra=
\sum_{q}\la\tsig_{uv},\tsig_{ij}(\tau)\ra^{(q)}e^{\rmi q(\om_{\rm p}-\om_{\rm d})t}$
in the stationary state,
and the static component ($q=0$) is measured in actual experiments. 
The power spectrum 
is then given by
\begin{widetext}
\beq
S(\om)=\hbar\om\times{\rm Re}\sum_{u,v,i,j}\frac{\sqrt{\tkap_{uv}\tkap_{ji}}}{\pi}
\int_0^{\infty}d\tau
e^{\rmi(\om-\om_{\rm d})\tau}\la\tsig_{uv},\tsig_{ij}(\tau)\ra^{(0)}.
\label{eq:ps}
\eeq
\end{widetext}

\begin{figure}[h]
\includegraphics[width=60mm]{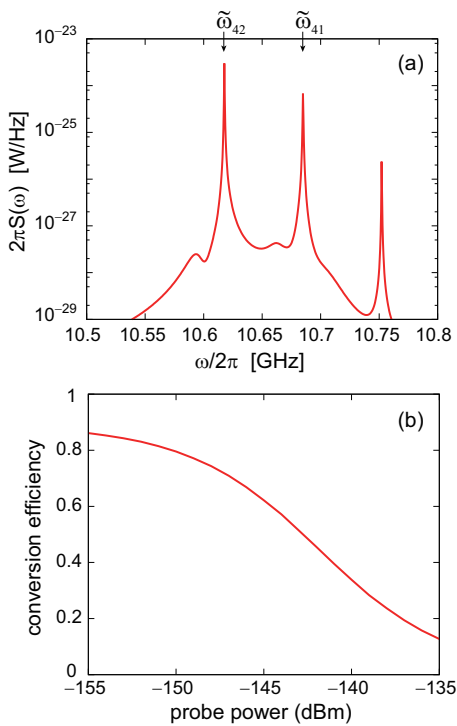}
\label{fig1s}
\caption{(a)~Power spectrum density of the reflected wave.
The input field is tuned to $\tom_{41}$,
whereas the principal peak of the spectrum apears at $\tom_{42}$. 
(b)~Down-conversion efficiency as a function of the probe power.}
\label{fig:S2}
\end{figure}

The power spectrum of the reflection field is shown in Fig.~\ref{fig:S2}(a), 
where the drive field is set at 
$\delta\om_{\rm d}/2\pi=-64$~MHz and $P_{\rm d}=P_{\rm d4}$, and 
the probe field is set at $\om_{\rm p}=\tom_{41}$
and $P_{\rm p}=-146.2$~dBm
[the upper dip of Fig.~3(a) of the main text].
We observe that the principal peak of the output power spectrum 
appears at the down-converted frequency $\tom_{42}$.
We define the down-conversion efficiency $\eta$ with the area of 
the down-converted peak normalized by the input flux,
i.e., $\eta=\int_{\om'}^{\om''} {\rm d}\om S(\om)/\hbar\om|E_{\rm p}|^2$,
where we have chosen $\om',\om''$=$\tom_{42}\pm 2\pi\times 30$~MHz.
In Fig.~\ref{fig:S2}(b), 
the conversion efficiency $\eta$ is plotted as a function 
of the input probe power $P_{\rm p}$.
Here, fixing the drive frequency at $\delta\om_{\rm d}/2\pi=-64$~MHz
and varying the probe power $P_{\rm p}$,
we performed the same simulation as Fig.~3(c) of the main text
to identify the dip position $(P_{\rm d4}, \tom_{41})$.
Then, fixing the drive power at $P_{\rm d}=P_{\rm d4}$ 
and the probe frequency at $\om_{\rm p}=\tom_{41}$,
we calculated the power spectrum by Eq.~(\ref{eq:ps})
and integrated the peak centered at $\tom_{42}$.
We observe that $\eta$ is a decreasing function of $P_{\rm p}$, 
which is due to population of $|\ttwo\ra$ due to continuous driving by the probe.
The conversion efficiency amounts to 67.7\% at $P_{\rm p}=-146.2$~dBm,
which roughly reproduces the measured efficiency of 74$\pm$8\%. 


\section{Calibration of input microwave power}
Owing to the nonlinear inductance of a Josephson junction embedded in a coplanar waveguide (CPW) resonator, a resonant frequency $\omega_{\rm r}$ is dependent on an input-probe power $P_{\rm p}$~\cite{Siddiqi04}. We measure $\omega_{\rm r}$ as a function of $P_{\rm p}$ and fit it by a theoretical model to precisely determine the microwave power input to the device. In this section, we explain the theoretical model and show the result of the numerical fitting. 

\begin{figure}[t]
\includegraphics[width=8.0cm]{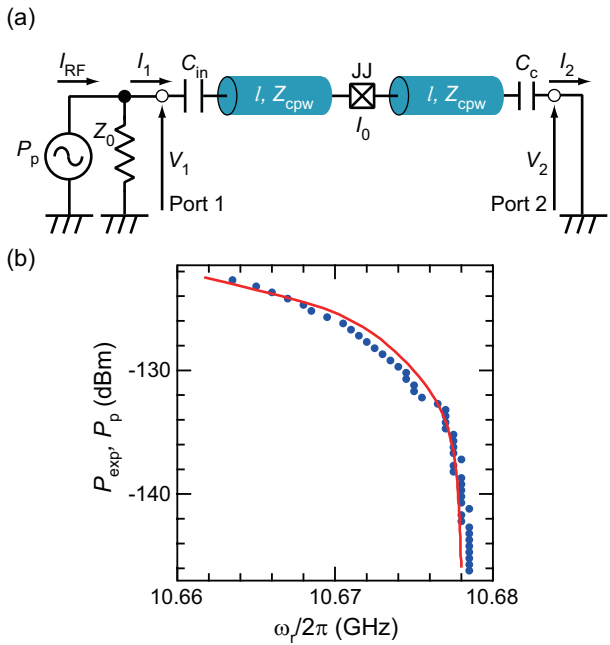}
\label{PCal}
\caption{(a) Schematic of a CPW resonator with a Josephson junction. (b) Input power dependence of the resonant frequency $\omega_{\rm r}$ of the resonator. The dots show the measured $\omega_{\rm r}$ plotted as a function of $P_{\rm exp}$, while the line is the theoretical fitting for $\omega_{\rm r}$ drawn as a function of $P_{\rm p}$.}
\label{fig:S3}
\end{figure}

Figure~\ref{fig:S3}(a) shows a schematic of a $\lambda /2$ CPW resonator which is used to read out the qubit states in the main text. The resonator is made of a 50-nm-thick Nb film sputtered on an oxidized high-resistivity silicon wafer. Width and a length of a center conductor of the resonator are $a=10~\mu$m and $2l=4.3$~mm, respectively, and a gap between the center conductor and a ground plane is $b=5.8~\mu$m. An Al Josephson junction with a designed critical current of $I_0=0.7~\mu$A was fabricated at the middle of the center conductor. 
Coupling capacitances between the microwave feedline and the resonator $C_{\rm in}$ and between the resonator and the qubit $C_{\rm c}$ are 15~fF and 4~fF, respectively~\cite{Inomata12}.

By introducing the amplitude $\Delta$ of the superconducting-phase oscillation, an inductance of the Josephson junction and the current $I_{\rm J}$ across the junction are given by~\cite{Watanabe09}
\begin{equation}\label{LJ}
L_{\rm J}(\Delta)=\frac{\Delta}{2J_1(\Delta)}L_{\rm J0},
\end{equation}
\begin{equation}\label{IJ}
I_{\rm J}/I_{\rm 0}=2J_1(\Delta)\sin \omega t,
\end{equation}
where $J_1$ is the Bessel function of the first kind. Note that at the limit of $\Delta = 0$,  $L_{\rm J}$ is equal to $L_{\rm J0} = \Phi_0 I_0/2\pi$ which is a (linear) Josephson inductance.
We characterize the circuit by using the transmission ($ABCD$) matrices~\cite{Pozer}. The matrix $T_{\rm Z}$ for an impedance $Z$ is
\begin{equation}\label{Zmat}
T_{\rm Z}=
\left(
\begin{array}{cc}
1 & Z\\
0 & 1\\
\end{array}
\right),
\end{equation}
and that for a section of lossless CPW with a length $l$ is
\begin{equation}\label{CPWmat}
T_{\rm CPW}=
\left(
\begin{array}{cc}
\cos \beta l & jZ_{\rm CPW}\sin \beta l \\
j(Z_{\rm CPW})^{-1}\sin \beta l & \cos \beta l \\
\end{array}
\right),
\end{equation}
where $j$ is the imaginary unit, $Z_{\rm CPW}$ is the characteristic impedance of the CPW, $\beta=\omega /v_{\rm p}$, and $v_{\rm p}$ is the phase velocity. Therefore, the matrix for the resonator (between the port 1 to 2 in Fig.~\ref{fig:S3}(a)) is given by
\begin{equation}\label{ABCmat}
\left(
\begin{array}{cc}
A & B \\
C & D \\
\end{array}
\right)
=T_{\rm C1}T_{\rm CPW}T_{\rm JJ}T_{\rm CPW}T_{\rm C2},
\end{equation}
where $T_{\rm C1}$, $T_{\rm JJ}$, and $T_{\rm C2}$ are the matrices for $C_{\rm in}$, the junction, and $C_{\rm c}$. They are obtained by replacing $Z$ in Eq.~(\ref{Zmat}) by their impedance of $(j\omega C_{\rm in})^{-1}$, $[(j\omega L_{\rm J})^{-1}+j\omega C_{\rm J}]^{-1}$, and $(j\omega C_{\rm c})^{-1}$, respectively, where $C_{\rm J}$ is a junction capacitance estimated to be 8.5~fF based on the junction size.

To simulate $P_{\rm p}$ dependence of $\omega_{\rm r}$, the port~1 of the resonator is connected to a microwave source which can expressed as an ideal current source and $Z_0=50~\Omega$ in parallel, and the port 2 is shorted [see Fig.~\ref{fig:S3}(a)].
Therefore, the transport equation from the port 1 to the port 2 is written as
\begin{equation}\label{TrEq}
\left(
\begin{array}{c}
V_1 \\
I_1 \\
\end{array}
\right)
=
\left(
\begin{array}{cc}
A & B \\
C & D \\
\end{array}
\right)
\left(
\begin{array}{c}
V_2 \\
I_2 \\
\end{array}
\right).
\end{equation}
Note $V_2=0$ because the port 2 is shorted to the ground. 
Using the output from the ideal current source $I_{\rm RF}=I_1+V_1/Z_0$, $P_{\rm p}$ in dBm unit is written as
\begin{equation}\label{probe}
P_{\rm p}=10\log \left( \frac{Z_0 I_{\rm RF}^2}{8\times 10^{-3}} \right).
\end{equation}
From the transmission matrix, we obtain the scattering matrix element $S_{11}$ or reflection coefficient as
\begin{equation}\label{Spara}
S_{11}=\frac{A+B/Z_0-CZ_0-D}{A+B/Z_0+CZ_0+D}.
\end{equation}
Thus, for a fixed $\Delta$, we can calculate $P_{\rm p}$ from Eq.(\ref{probe}) and $\omega_{\rm r}$ from Eq.(\ref{Spara}).
In the fitting, we used $x$, $I_0$, and $Z_{\rm CPW}$ as fitting parameters. Here, $x$ is a constant to account for the difference between $P_{\rm p}$ and $P_{\rm exp}$, namely, $P_{\rm p} = x P_{\rm exp}$, where $P_{\rm exp}$ is the probe microwave power estimated from total losses and attenuations in the input line.
$Z_{\rm CPW}$ can deviate from the designed value of 50 $\Omega$ due to the kinetic inductance of the Nb thin film. 

Figure~\ref{fig:S3}(b) shows the result of the fitting. The experimental data is well fitted by the theoretical model. We obtained $x=0.998$, $I_0=0.689~\mu$A, and $Z_{\rm CPW}=52.1~\Omega$ as fitting parameters.
$I_0$ agrees well with the designed value of $0.7~\mu$A.
Also, $v_{\rm p}$ estimated from $Z_{\rm CPW}$ and the capacitance of the resonator~\cite{Wen69} is consistent with the value in our previous work~\cite{Inomata09}.

\bibliography{main_text}

\begin{thebibliography}{29}
\bibitem [{JPA()}]{JPA_Note}%
  \BibitemOpen
  \href@noop {} {}\bibinfo {note} {The measurement with the qubit drive off
  calibrates the JPA gain. To compensate fluctuations of the JPA gain, we
  determine the gain in every measurement}\BibitemShut {NoStop}%
\end{thebibliography}

\end{document}